\documentstyle[12pt]{article}
\begin{document}

\begin{center}
{\Large {\bf
THE $N^*(1520) \rightarrow \Delta \pi$ AMPLITUDES EXTRACTED FROM THE 
$\gamma p \rightarrow \pi^+ \pi^- p$ REACTION AND COMPARISON TO QUARK MODELS
}}
\end{center}

\vspace{1cm}

\begin{center}
J.A. G\'omez Tejedor, F. Cano and E. Oset
\end{center}

\begin{center}
Departamento de F\'{\i}sica Te\'orica and IFIC,\\
Centro Mixto Universidad de Valencia-CSIC\\
46100 Burjassot (Valencia), Spain
\end{center}

\vspace{3cm}

\begin{abstract}
{\small{
The $\gamma p \rightarrow \pi^+ \pi^- p$ reaction, in combination with data
from the $\pi N \rightarrow \pi \pi N$ reaction, allows one to obtain the $s$-
and $d$-wave amplitudes for the $N^*(1520)$ decay into $\Delta \pi$ with
absolute sign with respect to the $N^*(1520) \rightarrow N \gamma$ helicity
amplitudes.
In addition one obtains the novel information on the $q$ dependence of the
amplitudes.
This dependence fits exactly with the predictions of the
non-relativistic constituent quark models. The absolute values provided by
these models agree only qualitatively, and a discussion is done on the
reasons for it and possible ways to improve.
}}
\end{abstract}

\vfill
{\it PACS}: 13.60.Le \hspace{.5cm}  14.20.Gk  \\

 \hfill

\vfill
\hfill
 {\it Preprint} {\bf FTUV/95-54, IFIC/95-56}
\vfill
\hfill
{\it Preprint} {\bf nucl-th/9510007}

\newpage

A recent detailed study of the $\gamma p \rightarrow \pi^+ \pi^- p$
reaction \cite{yo}, improving on the model of L\"uke and S\"oding \cite{luke},
together with a new 
experiment \cite{daphne,bragieri}, has stressed the role
of the $N^*(1520)$ resonance which is essential to understand the total
cross section for the $\gamma p \rightarrow \pi^+ \pi^- p$ reaction for photon
energies around $E_{\gamma}=700$ $MeV$.

In ref. \cite{yo} it is shown that the peak observed in the total cross
section for the $\gamma p \rightarrow \pi^+ \pi^- p$ reaction around
$E_{\gamma}=700$ $MeV$ \cite{daphne,bragieri,experiment,gialanella} is
due to the interference
of the dominant term of the reaction, the contact gauge term
$\gamma N \rightarrow \Delta \pi$ (the $\Delta$-Kroll-Ruderman term) and
the $\gamma N \rightarrow N^*(1520) \rightarrow \Delta \pi$ process, when
the decay of the $N^*(1520)$ into $\Delta \pi$ is through the $s$-wave.
More recently, these results have been confirmed in ref. \cite{laget}
where a simplified model with respect to the one in ref. \cite{yo} is used
for different isospin channels of the $\gamma p \rightarrow \pi \pi N$
reaction.

In this paper we show how we can obtain the amplitudes for the $N^*(1520)
\rightarrow \Delta \pi$ process 
from the $\gamma p \rightarrow \pi^+ \pi^- p$
reaction and their momentum dependence, which provides a nice test for the
quark models.

The first ingredient in the $\gamma N \rightarrow N^*(1520) \rightarrow
\Delta \pi$ process is the $N^*(1520) N \gamma$ coupling,
which is given by \cite{yo}:

\begin{equation}                \label{hn'nf}
-i \delta H_{N'^* N \gamma} =
i \tilde{g}_{\gamma} \vec{S} \cdot \vec{\varepsilon} \: + \:
\tilde{g}_{\sigma} \left( \vec{\sigma} \times \vec{S} \, \right)
\cdot \vec{\varepsilon}
\end{equation}
\noindent
by means of which one reproduces the two helicity decay amplitudes. In Eq.
(\ref{hn'nf}) $\vec\sigma $ are the ordinary spin Pauli matrices,
$\vec S$ is the transition spin operator from $1/2$ to $3/2$ and $\vec
\epsilon$ the photon polarization vector in the Coulomb Gauge.
From the average experimental values of the helicity amplitudes  given in
\cite{pdg94} we get $\tilde{g}_{\gamma}=0.108$ and
$\tilde{g}_{\sigma}= -0.049.$

For the $N^*(1520) \Delta \pi$ coupling, the simplest Lagrangian allowed by
conservation laws is given by \cite{yo}:
\begin{equation}                    \label{simple2}
{\cal L}_{N'^* \Delta \pi} =
i \tilde{f}_{N'^* \Delta \pi} \overline{\Psi}_{N'^*}  \phi^{\lambda}
  T^{\lambda} \Psi_{\Delta}
\: + \: h.c.
\end{equation}
\noindent
where ${\Psi}_{N'^*}$, $\phi^{\lambda}$ and $\Psi_{\Delta}$ stand for the
$N^*(1520)$, pion and $\Delta(1232)$ field respectively,
$\, T^{\lambda}$ is the
$1/2$ to $3/2$ isospin transition operator.

However, such a Lagrangian only gives rise to $s$-wave $N^*(1520) \rightarrow
\Delta \pi$
decay, while experimentally we know that there is  a large fraction
of decay into $d$-wave too \cite{pdg94,manley}. Furthermore, the
amplitude of Eq. (\ref{simple2}) provides a spin independent
amplitude, while non relativistic constituent quark models (NRCQM) give
a clear spin dependence in the amplitude.
We propose here for this coupling the following Lagrangian, which,
as we shall see, is supported by both the experiment and the NRCQM.
The Lagrangian is given by
\begin{equation}                \label{ln'dp}
{\cal L}_{N'^* \Delta \pi} =
i  \overline{\Psi}_{N'^*}
\left(
   \tilde{f}_{N'^* \Delta \pi} \: - \:
   \frac{\tilde{g}_{N'^* \Delta \pi}}{{\mu}^2}
   S_i^{\dagger} \partial_i \, S_j \partial_j
\right)
\phi^{\lambda} T^{\lambda} {\Psi}_{\Delta}
\: + \: h.c.
\end{equation}
\noindent
with $\mu$ the pion mass.

This Lagrangian gives us the vertex contribution to the $N^*(1520)$ decay
into $\Delta \pi$:
\begin{equation}                \label{hn'dp}
- i \delta H_{N'^* \Delta \pi} = -
\left(
  \tilde{f}_{N'^* \Delta \pi} \: + \:
  \frac{\tilde{g}_{N'^* \Delta \pi}}{\mu^2}
  \vec{S}^{\dagger} \cdot \vec{q}  \, \vec{S} \cdot \vec{q}
\right)
T^{\lambda}
\end{equation}
\noindent
where $\vec{q}$ is the pion momentum.
In order to fit the coupling constants $\tilde{f}_{N'^* \Delta \pi}$
and $\tilde{g}_{N'^* \Delta \pi}$ to the
experimental amplitudes in $s$- and
$d$-wave \cite{pdg94} we make a partial wave expansion \cite{pin} of the
transition amplitude $N^*(1520)$ to $\Delta \pi$ from a state of spin 3/2
and third component $M$, to a state of spin $3/2$ and third component
$M'$, following the standard ``baryon-first'' phase convention \cite{phase}:
$$
- i \, \langle \frac{3}{2} M' | \delta H_{N'^* \Delta \pi}
                              | \frac{3}{2}, M \rangle =
A_s \, Y_0^{M-M'} (\theta, \phi) \: + \:
$$
\begin{equation}                       \label{pwa}
A_d \, \langle 2, \frac{3}{2}, M-M', M' |
    2, \frac{3}{2}, \frac{3}{2}, M \rangle \, Y_2^{M-M'}(\theta , \phi)
\end{equation}
\noindent
where $\langle j_1 j_2 m_1 m_2 | j_1 j_2 J M \rangle$ is the corresponding
Clebsch-Gordan coefficient, $Y_l^m(\theta, \phi)$ are the spherical
harmonics, and $A_s$ and $A_d$ are the $s$- and $d$-wave partial amplitudes
for the $N^*(1520)$ decay into $\Delta(1232)$ and $\pi$, which are given by:
\begin{equation}            \label{AsAd}
\begin{array}{l}
A_s = - \sqrt{4 \pi}
     \left( \tilde{f}_{N'^* \Delta \pi} \: + \:
            \frac{1}{3} \tilde{g}_{N'^* \Delta \pi} \frac{\vec{q}\,^2}{\mu^2}
     \right) \\ \\
A_d = \frac{\sqrt{4 \pi}}{3} \tilde{g}_{N'^* \Delta \pi} \frac{\vec{q}\,^2}{\mu^2}
\end{array}
\end{equation}

   From Eq. (\ref{pwa}) we get the expression for the $N^*(1520)$ decay width
into $\Delta \pi$:
\begin{equation}        \label{gammadp}
\Gamma =
 \frac{1}{4 \pi^2} \, \frac{m_{\Delta}}{m_{N'^*}} \, q
 \left( |A_s|^2 + |A_d|^2 \right) 
\theta (m_{N'^*} - m_\Delta - \mu )
\end{equation}
\noindent
where $q$ is the momentum of the pion. We then fit the $s$- and $d$-wave
parts of $\Gamma$ to the average
experimental values \cite{pdg94} by keeping the
ratio $A_s/A_d$ positive as deduced from the experimental analysis of the
$\pi N \rightarrow \pi \pi N$ reaction \cite{manley}. We get then
two different solutions which differ only
in a global sign,
\begin{equation}                   \label{2sol}
\begin{array}{rll}
{\it (a)} & \hspace{1cm} \tilde{f}_{N'^* \Delta \pi} = 0.911  &
            \hspace{1cm} \tilde{g}_{N'^* \Delta \pi} =-0.552 \\
{\it (b)} & \hspace{1cm} \tilde{f}_{N'^* \Delta \pi} =-0.911 &
            \hspace{1cm} \tilde{g}_{N'^* \Delta \pi} = 0.552 \\
\end{array}
\end{equation}

Now, the $\gamma p \rightarrow \pi^+ \pi^- p$ reaction allows us to
distinguish between both solutions, hence providing the relative sign with
respect to the $N^*(1520) \rightarrow \gamma N$ amplitude.

In Fig. 1 we have plotted the total cross section for both solutions
(solid lines).
As we can see, only solution ${\it (a)}$ fits the experiment, while the other
one under-estimates the experimental cross section by a large amount.
In Fig. 1 we also show the uncertainties in the cross section due to the
experimental errors in the $N^*(1520)$ helicity amplitudes and $s$- and
$d$-wave $\Delta \pi$ decay widths (region between dashed lines).
These errors are calculated by evaluating the results a large number of
times, $N$, with random values of the couplings within experimental errors.
The deviation $\sigma$ from the mean, $\overline{x}$, is then obtained as
\cite{bevington}:
\begin{equation}
\sigma^2 =\frac{\sum_i (x_i - \overline{x} )^2}{N-1}
\end{equation}

For the width of the $N^*(1520)$ in the propagator we have taken the
explicit decay into the dominant channels ($N \pi$, $\Delta \pi$, $N \rho$)
with their energy dependence, improving on the results of \cite{yo} where the
energy dependence was associated to the $N \pi$ channel.

Because of the $N^*(1520)$ is a $d$-wave resonance, the energy dependence
of the decay width into $N \pi$ is given by:

\begin{equation}
\Gamma_{N'^* \rightarrow N \pi} (\sqrt{s}) =
\Gamma_{N'^* \rightarrow N \pi} (m_{N'^*})
\frac{q^5_{c.m.}(\sqrt{s})}{q^5_{c.m.}(m_{N'^*})}
\, \theta (\sqrt{s} - m -\mu)
\end{equation}

\noindent
where
$\Gamma_{N'^* \rightarrow N \pi} (m_{N'^*})= 66$ $MeV$ \cite{pdg94},
$q_{c.m.}(m_{N'^*})=456$ $MeV$ and
$q_{c.m.}(\sqrt{s})$ is the momentum of the decay pion in the $N^*(1520)$
rest frame.

For the $\Delta \pi$ channel, the energy dependence of the decay width 
is given by Eq. (\ref{gammadp}).

Finally, for the $N^*(1520)$ decay into $N \pi \pi$ through the $N \rho$
channel is given by:

\begin{equation}
\Gamma_{N'^* \rightarrow N \rho [\pi \pi]} =
\frac{m}{6(2 \pi)^3}
\frac{m_{N'^*}}{\sqrt s}
g_\rho^2  f_\rho^2
\int d \omega_1 d \omega_2 
| D_\rho(q_1 + q_2) | ^2
( \vec q_1 - \vec q_2 \, )^2
\end{equation}
where $q_i=(\omega_i ,\vec q_i)$ ($i=1,2$) are the fourmomenta of the
outgoing pions,
$  D_\rho(q_1 + q_2) $ is the $\rho$ propagator including the $\rho$ width,
$f_\rho$ is the $\rho \pi
\pi$ coupling constant ($f_\rho = 6.14$), and $g_\rho$ is the $N'^* N \rho$
coupling constant ($g_\rho=7.73$)
that we fit from the experimental $N'^* \rightarrow N \rho [\pi \pi]$ decay
width \cite{pdg94}.
A slightly different, although equivalent  treatment can be found in ref.
\cite{laget}.

The differences
induced in the cross section from these improvements with respect to
ref. \cite{yo} are, however, very small.

In Fig. 1 we are also plotting the experimental results of ref. \cite{daphne}
with the DAPHNE acceptance, together with our theoretical results with this
acceptance (long dashed line). This is proper to do since the experimental
total cross section is extrapolated from the measured one using the model
of ref. \cite{laget}.

We have also checked possible effects coming from off-shell effects in the
propagators and vertices of the spin $3/2$ particles ($\Delta$ and
$N^*(1520)$) \cite{arndt,benmerrouche}. By taking $A=-1$ and $Z \in [-1/2,1/2]$
the changes observed in the cross section are of the order of $1$\%.

We  should note that the interference between the
$\gamma N \Delta \pi$-Kroll-Ruderman and the $\gamma N \rightarrow
N^*(1520) \rightarrow \Delta \pi$ terms changes sign around
$\sqrt{s}=m_{N'^*}$ where the real part of the $N^*(1520)$ propagator
changes sign. This means that the on-shell 
value of the amplitudes $A_s$ and $A_d$
for the $N^*(1520) \rightarrow \Delta \pi$ decay 
plays no role at this energy and what matters is
the value of $A_s$ (the one that interferes) at values of $q$ other 
than the one from the decay of the $N^*(1520)$ on-shell.
This brings us to the $q$ dependence of the amplitude. 
While the $A_d$ part should have the $q^2$ dependence exhibited
in Eq. (\ref{AsAd}), the combination of $q^2$ which appears in $A_s$ is
given by the chosen Lagrangian. One could, however,
postulate other Lagrangians
which would lead to a different combination.
In order to investigate the most general $q^2$ dependence of $A_s$ we
substitute $\tilde{f}_{N'^* \Delta \pi}$ by

\begin{equation}                   \label{epsilon}
\tilde{f}_{N'^* \Delta \pi}
\left( 1 \: + \: \epsilon \, \frac{\vec{q}\,^2 - \vec{q}\,^2_{on-shell}}{\mu^2}
\right)
\end{equation}
\noindent
where $\vec{q}$ is the momentum of the decay pion, and
$\vec{q}\,^2_{on-shell}$ is de momentum of the pion for a on-shell $N^*(1520)$
decaying into $\Delta \pi$ ($|\vec{q}_{on-shell}|$ = $228 \: MeV$), and then
we change $\epsilon$ comparing the results to the data. We find that, to a
good
approximation, $\epsilon=0$ gives the best agreement with the data, hence
supporting the Lagrangian of Eq. (\ref{ln'dp}).

In a recent paper \cite{hepph} we use the information obtained here,
together with all the other needed effective Lagrangians, in order to study
the $\gamma N \rightarrow \pi \pi N$ reaction in all the isospin channels.

Next we pass to see what the NRCQM have to say with respect to this novel
information. We followed a model designed by Bhaduri {\it et al.} 
\cite{BHADURI81} to describe the mesonic spectrum and which was used later 
on by Silvestre-Brac {\it et al.} \cite{SILVESTRE86} in the baryonic sector.
The model has as starting point the quark-quark ($qq$) potential 

\begin{equation}
\label{potencialdebhaduri}
V_{qq}= \frac{1}{2} \sum_{i<j} 
                \left( -\frac{\kappa}{r_{ij}} + \frac{r_{ij}}{a^{2}} 
                + \frac{\kappa_{\sigma}}{m_{i} m_{j}} 
                \frac{\exp{(-r_{ij}/r_{0}})}{r_{0}^{2} r_{ij}}
                \vec{\sigma_{i}} \cdot \vec{\sigma_{j}} -D \right)
\end{equation}

\noindent
incorporating the basic QCD motivated confining, coulombic and
spin-spin $qq$ interactions, where the parameters are chosen in order to
reproduce the low energy baryonic spectrum.

        In order to study strong pionic decays $B \rightarrow B' \pi$ we
shall follow the elementary emission model (EEM) in which the decay takes
place through the emission of a (point-like) pion by one of the quark.
Some choices for
the $qq\pi$ Hamiltonian are possible. We quote here a pseudovector interaction

\begin{equation}
H_{qq\pi} = \frac{f_{qq\pi}}{\mu} \overline{\Psi}_{q}(x)
\gamma^{\nu} \gamma_{5}
        \vec{\tau} \Psi_{q}(x) \partial_{\nu} \vec{\phi}(x)
\label{pseudovector}
\end{equation}

        The non-relativistic approach comes from the non-relativistic
expansion of Eq. (\ref{pseudovector}) 
in powers of $(p/m_q)$, where $p$ is the quark momentum operator.
Up to first order in $(p/m_q)$ the Hamiltonian governing the transition
$B \rightarrow B' \pi^{\alpha}$ has this form:

\begin{equation}                         \label{Hqqpi}
H_{qq\pi} \propto f_{qq\pi} \, (\tau_{\alpha})^{\mbox {\tiny \dag}}
\left[ \vec{\sigma} \cdot \vec{q} e^{- i \vec{q} \vec{r}}  -
\frac{\omega_{\pi}}{2 m_{q}}\vec{\sigma} \cdot 
\; (\vec{p} e^{- i \vec{q} \vec{r}} +
e^{- i \vec{q} \vec{r}} \vec{p} \,) \right]
\label{structureoperator}
\end{equation}
 
        The isospin ($\vec{\tau}$), spin ($\vec{\sigma}$) and momentum
($\vec{p}\,$) 
operators stand for the quark responsible for the emission, and
$\omega_{\pi}$, $\vec{q}$
are the energy and momentum of the emitted pion respectively.
In Eq. 
(\ref{structureoperator}) one distinguishes the term proportional to $\vec{q}$
(direct term) and the recoil term with the $\vec{p}$ structure.

        There are two independent helicity amplitudes for the 
$N^{*}(1520) \rightarrow 
\Delta \pi$ decay. If we take the quantization axis along
the pion momentum in the resonance rest frame, the 
helicity amplitudes correspond
to a resonance spin projection, and we denote them as $A_{1/2}$
and $A_{3/2}$.
After performing the calculations the ratio between them is:

\begin{equation}
\frac{A_{3/2}}{A_{1/2}} = \frac{-C_{\mbox {\scriptsize REC}}}
{C_{\mbox {\scriptsize DIR}} - C_{\mbox {\scriptsize REC}}}
\label{ratio}
\end{equation}

\noindent
where $C_{\mbox {\scriptsize DIR (REC)}}$ is the contribution from the
direct (recoil) term. Rigorously, what we call $C_{\mbox {\scriptsize DIR}}$
contains a small piece, proportional to $\frac{\omega_{\pi}}{6 m_{q}}$ coming
from the recoil term in (\ref{structureoperator}).

The amplitudes $A_s$ and $A_d$ of Eq. (\ref{AsAd}) can be expressed in terms of
these helicity amplitudes as follows \cite{martin}:
\begin{equation}
\begin{array}{l}
A_{d}  \propto A_{3/2} - A_{1/2} \\ \\
A_{s} \propto A_{3/2} + A_{1/2}
\end{array}
\end{equation}

\noindent
and their ratio in the EEM is given by

\begin{equation}
\frac{A_{d}}{A_{s}} =
\frac{C_{\mbox {\scriptsize DIR}}}{2 C_{\mbox {\scriptsize REC}} -
C_{\mbox {\scriptsize DIR}}} = + 0.156
\end{equation}

\noindent
where we have quoted the value obtained with the Bhaduri potential 
(\ref{potencialdebhaduri}). The experimental value for this ratio is
$1.2$ \cite{pdg94}.

        Let us first discuss the sign. Notice that if only the direct term 
were present, the relative sign would be negative (the so-called SU(6)$_{W}$
signs) \cite{LEYAOUANC88}. 
The introduction of the first order (recoil) contribution provokes a
change of sign (the anti-SU(6)$_{W}$ situation) in agreement with the
experiments \cite{manley}. This fact was pointed out long ago by Le 
Yaouanc {\it et al.} \cite{LEYAOUANC75} by using the $^{3}P_{0}$ model
(that could be regarded to some extent as a $(p/m_q)$ model). We have checked
this sign with a wide variety of $qq$ potentials and with the $^{3}P_{0}$
model also, and the anti-SU(6)$_{W}$ signs remain. Moreover, we have explored
in the EEM with harmonic oscillator wave functions under which conditions are
the SU(6)$_{W}$ signs recovered. The answer is that the radius of the nucleon
has to be larger than $\approx$ 1 fm. Certainly, spectroscopy does not support
such a big quark core radius. Hence, as a quite model independent conclusion,
we can say that the 
recoil term is crucial to explain the anti-SU(6)$_{W}$ signs, and it is 
generally bigger than the direct term. 

        It is interesting to contrast the model prediction with the information
on the $q$ dependence which our analysis of the experiment has provided
for $A_{s}$ and $A_{d}$. From Eq. (\ref{AsAd}) we find

\begin{equation}             \label{AsAd2}
\begin{array}{l}
A_s + A_d = - \sqrt{4 \pi}
     \tilde{f}_{N'^* \Delta \pi} \\ \\
A_d = \frac{\sqrt{4 \pi}}{3} \tilde{g}_{N'^* \Delta \pi} 
\frac{\vec{q}\,^2}{\mu^2}
\end{array}
\end{equation}

        Equation (\ref{AsAd2}) summarizes in a practical way the empirical 
$q$ dependence of the amplitudes. Now let us see what the NRCQM gives. Eqs.
(\ref{AsAd2}) can be recast in terms of the helicity amplitudes as

\begin{equation} 
\begin{array}{l}
A_{3/2} \propto \tilde{f}_{N'^* \Delta \pi}  \\ \\
A_{3/2} - A_{1/2} \propto \tilde{g}_{N'^* \Delta \pi} \vec{q}\,^2
\end{array}
\end{equation}

\noindent 
which in terms of Eq. (\ref{ratio}), by means of the direct and recoil terms
of Eq. (\ref{structureoperator}), can be rewritten  as

\begin{equation}
\begin{array}{l}
C_{\mbox {\scriptsize REC}} \propto \tilde{f}_{N'^* \Delta \pi} \\ \\
C_{\mbox {\scriptsize DIR}} \propto \tilde{g}_{N'^* \Delta \pi}
\vec{q}\,^2  
\end{array}
\end{equation}

        Now it is straightforward to see that this is indeed the case.
The $N^{*}(1520) \rightarrow \Delta \pi$ 
transition matrix element with the direct
term of Eq. (\ref{structureoperator}) requires the second term in the
expansion of $e^{-i \vec{q} \vec{r}}$, since $N^*(1520)$ contains a radial
excitation with
respect to the $\Delta(1232)$. Hence, the direct term is proportional to 
$\vec{q} \, ^{2}$. On the other hand the recoil term gets the dominant 
contribution from the unity in the expansion of the exponential and hence it is
momentum independent. Thus the quark model prediction for the $\vec{q}\,^{2}$
dependence of the amplitudes is in perfect agreement with experiment. However,
the strength of the terms and their ratio is not well reproduced. This is not 
surprising in view that the recoil term appears to be bigger than the 
direct one in the $(p/m_q)$ expansion of Eq. (\ref{Hqqpi}). The values met here
for $(p/m_q)$ are in
average bigger than 1 and one should then expect limitations due to the 
nonrelativistic character of the model.

The purpose of the present paper is not to solve this interesting problem
which has already caught attention of some groups \cite{KONIUK80,CAPSTICK94}.
Our purpose has been to show the novel experimental information about the
$q$ dependence of the $s$- and $d$-wave $\Delta \pi$ decay amplitudes of the
$N^*(1520)$ extracted from the $\gamma p \rightarrow \pi^+ \pi^- p$ reaction,
and how it fits with the structure of NRCQM. It also gives in addition an
absolute sign with respect to the $N^*(1520)$ helicity amplitudes which
agrees with the NRCQM.

As for the need to introduce relativistic effects to get the appropriate
strength of the $s$- and $d$-wave ratio it seems quite obvious, and some
results show that the ratio improves when this is done. The method of
\cite{KONIUK80} is probably an indirect way of introducing relativistic
effects by taking different factors in front of the two terms in Eq.
(\ref{Hqqpi}) which are the then fit to a large set of data. In ref.
\cite{CAPSTICK94} a $^3P_0$ model
with relativized hadronic wave functions
is used and contrasted to a
large set of hadronic decays of the baryon spectrum, and, concretely for the
$N^*(1520)$, the $s$- and $d$-wave $\Delta \pi$ decay ratio improves
considerably without still being in agreement with experimental data.
In Table I we show the results
obtained with all these models. These results
indicate the importance of the relativistic
effects and the need for more work.
Another possibility is explored in ref. \cite{CANO95} by making an expansion
in powers of $(p/E)$ instead of $(p/m_q)$.
However when trying to improve on this ratio it will be important to take
into account the new experimental constraint obtained in the present work,
and summarized in Eq. (\ref{AsAd2}). While the second equation, establishing
$A_d$ as a quadratic function of $q$, will come out relatively naturally in
most schemes, the independence of $q$ of $A_s + A_d$ of the first equation is
less than obvious and will pose a challenge to any new scheme.

In Fig. 1 we are also plotting the results obtained by using the strong and
electromagnetic couplings for the $N^*(1520)$ resonance from the work of refs.
\cite{CAPSTICK94,CAPSTICK92}.
The results obtained are very close to those obtained with our model.
This is so in spite that the individual electromagnetic and strong
couplings are in some disagreement with experiment \cite{pdg94}.
Indeed, the helicity amplitudes of \cite{CAPSTICK92} are smaller than
experiment and the $s$-wave $N^*(1520) \rightarrow \Delta \pi$ amplitude of
\cite{CAPSTICK94} (the relevant one in the interference) bigger tan the
experiment, and there is a certain compensation of both deficiencies in the
$\gamma p \rightarrow \pi^+ \pi^- p$ cross section. This observation is
interesting because it tells us that the fairness of a model for the $\gamma
N \rightarrow \pi \pi N$ reaction is not 
enough by itself and one has to contrast
the information provided by the model with the complementary experimental
information extracted from the $\pi N \rightarrow \gamma  N$ and the $\pi N
\rightarrow \pi \pi N$ reactions. As an example in ref. \cite{praga} we show
a model which gives equally good results as the present one in the $\gamma p
\rightarrow \pi^+ \pi^- p$ reaction and which has a ratio
$A_d/A_s$ of opposite sign to the experimental one.
These two examples show clearly the importance of using the information of
several experiments in order to obtain the proper information on the
properties of resonances, the $N^*(1520)$ in particular in the present case.

\vskip 2cm
This work has been partially supported by CICYT contract numbers AEN 93-1205
and AEN 93-0234. 
J.A.G.T. wishes to acknowledge financial support from the IVEI and F.C.
acknowledges the M.E.C. for a FPI fellowship.


\newpage

\section*{Figure Captions}
{\bf Fig. 1:} Continuous line: Total cross section for the $\gamma p \rightarrow \pi^+ \pi^- p$
             reaction for different solutions of
             $\tilde{f}_{N'^* \Delta \pi}$ and $\tilde{g}_{N'^* \Delta \pi}$
	     (see Eq. (\ref{2sol})).
Region between short-dashed lines: Uncertainties in the cross section
due to the
experimental errors in the $N'^*(1520)$ helicity amplitudes and $s$- and
$d$-waves $\Delta \pi$ decay widths.
Long-dashed line: Cross section integrated over the DAPHNE detector
acceptance \cite{daphne}.
Dash-dotted lines: Total cross section with the Capstick {\it et al.}
values of the strong and electromagnetic couplings
\cite{CAPSTICK94,CAPSTICK92}.

\section*{Table Captions}
{\bf Table 1:}$\Gamma_{d}/\Gamma_{s}$
Ratio for different models. Experimental
value from \cite{pdg94}.

\newpage
\begin{table}
\begin{center}
\begin{tabular}{@{}|l|c|c|c|c|@{}}
\hline \hline
 \rule{0pt}{4.5ex} & 
\begin{tabular}{c}
EEM  \\
with Eq. (\ref{Hqqpi}) 
\end{tabular} 
&
\begin{tabular}{c}
EEM  \\
\cite{KONIUK80} 
\end{tabular}
& 
\begin{tabular}{c} 
$^{3}P_{0}$  \\
\cite{CAPSTICK94}
\end{tabular} 
& Exp.\\[2ex]
\hline
\rule{0pt}{3.5ex}
$\frac{\Gamma_d}{\Gamma_{s}}$ & 0.024 & 0.139 & 0.069 & 1.4 $\pm$ 0.6
\\ [1.5ex]
\hline\hline    
\end{tabular}
\vskip 1cm
\end{center}
\end{table}

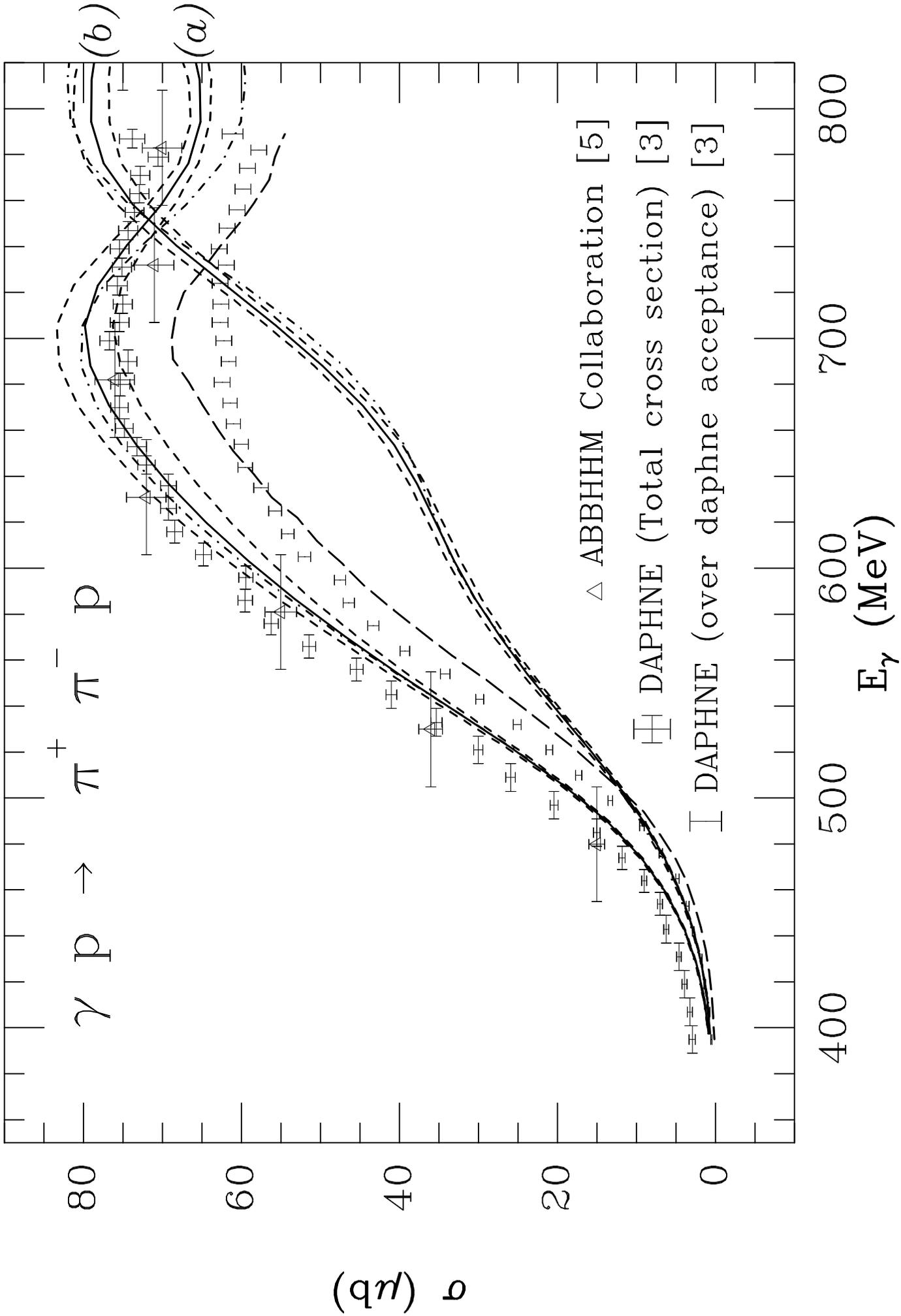
\end{document}